\documentclass{IEEElmag}
\usepackage[colorlinks,urlcolor=blue,linkcolor=blue,citecolor=blue]{hyperref}
\usepackage[hyphenbreaks]{breakurl}
\usepackage{physics}
\usepackage{color}
\usepackage{array}
\usepackage{siunitx}
\usepackage{graphicx}
\usepackage{CJKutf8}
\usepackage{amsmath}
\begin{document}

\sptitle{Spintronic Neuron Using a Magnetic Tunnel Junction for Low-Power Neuromorphic Computing}

\title{Spintronic Neuron Using a Magnetic Tunnel Junction for Low-Power Neuromorphic Computing}

\author{Steven Louis\affilmark{1}*} 
\author{Hannah Bradley.\affilmark{2}*}
\author{Cody Trevillian.\affilmark{2}*}
\author{Andrei Slavin.\affilmark{2}**}
\author{Vasyl Tyberkevych.\affilmark{2}*}

\affil{Department of Electrical and Computer Engineering, Oakland University, Rochester, MI 48309, USA}

\affil{Department of Physics, Oakland University, Rochester, MI 48309, USA}

%\IEEEmember{*Member, IEEE}
%\IEEEmember{**Fellow, IEEE}

\corresp{Corresponding author: Steven Louis (SLouis@oakland.edu).}

%\markboth{Preparation of Papers for \emph{IEEE Magnetics Letters}}{Author Name}

\begin{abstract}
This paper proposes a novel spiking artificial neuron design based on a combined spin valve/magnetic tunnel junction (SV/MTJ). 
Traditional hardware used in artificial intelligence and machine learning faces significant challenges related to high power consumption and scalability. 
To address these challenges, spintronic neurons, which can mimic biologically inspired neural behaviors, offer a promising solution. 
We present a model of an SV/MTJ-based neuron which uses technologies that have been successfully integrated with CMOS in commercially available applications. 
The operational dynamics of the neuron are derived analytically through the Landau-Lifshitz-Gilbert-Slonczewski (LLGS) equation, demonstrating its ability to replicate key spiking characteristics of biological neurons, such as response latency and refractive behavior. 
Simulation results indicate that the proposed neuron design can operate on a timescale of about 1 ns, without any bias current, and with power consumption as low as 50 \unit{\uW}.
\end{abstract}


\maketitle

\section{INTRODUCTION}\label{introduction}

Recent advancements in artificial intelligence (AI) and machine learning (ML) have been remarkable {[}Wu 2023{]}. 
However, traditional hardware used for AI applications often faces challenges such as high power consumption and significant costs {[}Strubell2020{]}. 
To mitigate these problems, recent research has proposed utilizing spintronic spiking neurons for AI and ML tasks {[}Khymyn 2018, Liu 2020, Louis 2022, Mitrofanova 2022, Ovcharov 2022, Bradley 2023, Rodirgues 2023, Manna 2023, Sengupta 2016, Cai 2019{]}. 
Spintronic spiking neurons are promising because they can exhibit biologically inspired features such as response latency, refractive behavior, and inhibition—all within a single component {[}Bradley 2023{]}. 
These characteristics have the potential to enhance the design of AI hardware.

In this article, we propose to use a combined spin valve (SV) and magnetic tunnel junction (MTJ) to form a biologically plausible artificial spiking neuron. 
Spin valves and MTJs have been extensively manufactured for commercial purposes. 
For over two decades, spin valves and MTJs were widely used as the read head for magnetic hard drives {[}Alzate 2015{]}.   
More recently, MTJs have been used in magnetoresistive random access memory (MRAM) as a non-volatile memory element. 
In MRAM, MTJs have been embedded in CMOS with a process length as small as 14 nanometers {[}Edelstein 2020{]}.
Given the longstanding practical implementation of SV/MTJs, developing a spintronic spiking neuron based on these technologies has the potential to be highly advantageous.

This paper presents a potential design of an SV/MTJ combination that can act as a spintronic spiking neuron. 
We demonstrate that the Landau-Lifshitz-Gilbert-Slonczewski (LLGS) equation for this particular configuration of SV/MTJ can be converted to the artificial neuron equation (ANE), which describes the dynamics of the artificial neuron {[}Bradley 2023{]}. 
In addition, this paper will present basic characteristics of the SV/MTJ neuron.

Overall, this proposed design of an SV/MTJ combination as a biologically plausible artificial spiking neuron has the potential to advance the field of AI and ML by providing a readily realizable hardware option for biologically realistic spintronic neurons.

\begin{figure}
\includegraphics{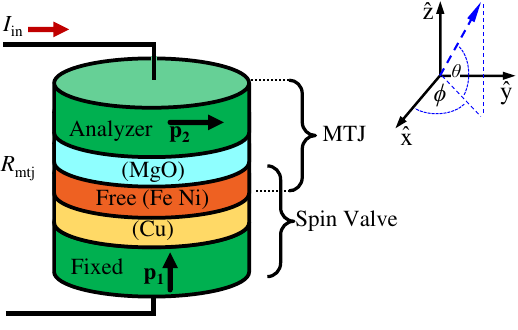}
\caption{\label{toon} Schematic of the SV/MTJ neuron.
The lower layers of the stack constitute a spin valve, while the upper layers constitute a magnetic tunnel junction. 
The input current flows through the stack as labeled, and $R_\text{mtj}$ is the resistance of the stack.}
\end{figure}

\section{NEURON DESIGN AND MATHEMATICAL MODEL}\label{design}

The proposed neuron is composed of a stack of ferromagnetic layers separated by spacer layers.
A sketch of the proposed stack is shown in Fig. \ref{toon}.
The bottom of the stack is a spin valve, consisting of a fixed layer, a copper spacer layer, and a free layer.
The fixed layer has a magnetic anisotropy that keeps its magnetization oriented out-of-plane (labeled $\vb{p}_1$ in the figure).
The top of the stack is a magnetic tunnel junction that extracts the signal.
This MTJ includes an analyzer layer with magnetization oriented in-plane (labeled $\vb{p}_2$ in the figure), and is separated from the free layer by an MgO spacer.

The SV/MTJ neuron operates as follows.
An electric current, $I_\text{in}$, flows through the stack and exerts a spin-transfer torque on the free layer magnetization. 
This electric current can be considered as an input to the neuron. 
The dynamics of the free layer magnetization under the influence of $I_\text{in}$ can be modeled in the macrospin approximation by the Landau–Lifshitz–Gilbert–Slonczewski (LLGS) equation {[}Slavin, 2009{]}:
\begin{equation}\dv{\vb{m}}{t} = |\gamma|\vb{B}_\text{eff} \times \vb{m} + \alpha_G \vb{m} \times \dv{\vb{m}}{t} +\sigma_j I_\text{in} \vb{m}\times\vb{m}\times\vb{p}_1   .  \label{llgs}\end{equation}
In this equation, $\vb{m}$ is the unit magnetization of the free layer, $\vb{p}_1$ is the unit magnetization of the polarizer, $\gamma$ is the gyromagnetic ratio, $\vb{B}_\text{eff}$ is the effective magnetic field, $\alpha_G$ is the Gilbert damping parameter, and $\sigma_j$ is the spin torque coefficient.
The effective field is $\vb{B}_\text{eff} = \vb{B}_\text{ext} + \vb{B}_\text{A} + \vb{B}_\text{d}$, where $\vb{B}_\text{ext}$ is an external magnetic field, $\vb{B}_\text{A}$ is the field due to magnetic anisotropy in the free layer, and $\vb{B}_d$ is the demagnetizing field.
The spin torque coefficient is given by $\sigma_j = |\gamma| \hbar \mu_0 \eta_0/(2 B_\text{d} e V)$, where $\hbar$ is the reduced Planck constant, $\eta_0$ is the spin polarization efficiency, $M_s$ is the free layer saturation magnetization, $e$ is the fundamental electric charge, and $V$ is the volume of the free layer.

The output of the neuron is the resistance $R_\text{mtj}$ across the SV/MTJ stack, which can be approximated by $R_\text{mtj} = R_0 + \Delta R (\vb{m}\cdot\vb{p}_2)$, where $\vb{p}_2$ is the unit magnetization of the analyzer layer, $\Delta R$ is the amplitude of MTJ resistance change, and $R_0$ is the average MTJ resistance.
A typical tunneling magnetoresistance ratio for an MTJ ranges from 200\% to 600\%, with a resistance on the order of 1 k$\Omega$ {[}Yuasa 2004, Parkin 2004{]}. 
While the spin valve exhibits giant magnetoresistance, the GMR ratio of the stack is typically between 5\% and 10\%, with a resistance on the order of 10 $\Omega$ {[}Baibich 1988, Parkin 2004{]}. 
Therefore, the contribution of the spin valve to $R_\text{mtj}$  is negligible compared to the magnetoresistance of the MTJ analyzer.

Please note that although there is magnetoresistance associated with the spin valve portion of the stack, the resistance is negligible compared to the magnetoresistance of the MTJ analyzer, as the contribution from the spin valve is typically several orders of magnitude smaller.

In the simulations that follow, we have chosen the polarizer magnetization to have an out-of-plane orientation ($\vb{p}_1 =\vu{z}$).
The analyzer layer magnetization ($\vb{p}_2$) was selected to have an in-plane orientation.
The free layer is considered isotropic ($\vb{B}_\text{A}= 0$) and is subject to a demagnetization field $\vb{B}_\text{d} = 1$~T in the $-\vu{z}$  direction, as well as an external field $\vb{B}_\text{ext}$ with a magnitude of 3 mT in the $\vu{x}$ direction.
This configuration resembles one that was previously fabricated and characterized in an experimental study {[}Houssameddine, 2007{]}.
It was chosen because the free layer magnetization is in-plane at equilibrium while exhibiting out-of-plane precession.
With this setup, the derivation of the ANE can be clearly demonstrated.
%In addition, other geometries of MTJs can also serve as spintronic neurons, as shown in simulations by {[}Rodrigues, 2023{]}.
%The following sections aim to inspire further research within the broader spintronics and MRAM community to develop neuromorphic hardware based on spiking neurons that can be integrated with CMOS and other conventional computing platforms.
Other MTJ geometries can also serve as spintronic neurons, as shown in simulations by {[}Rodrigues, 2023{]}. 
The following sections aim to inspire further research in the spintronics and MRAM community to develop neuromorphic hardware based on spiking neurons that can integrate with CMOS and other conventional computing platforms.

It should be noted that the spin-transfer torque from the analyzer layer is not included in (\ref{llgs}) due to the in-plane orientation of $\vb{p}_2$, as its contribution to the spin-transfer torque is, in general, negligible compared to $\vb{p}_1$. 
We have included a section in the supplementary material to further address minor quantitative changes due to $\vb{p}_2$.

Additional simulation parameters include $\gamma = 2 \pi \times 28 $~GHz/T,
$\alpha_G = 0.1$,
$\eta_0 = 0.35$,
$V = 3\times 10^4$~nm$^3$,
$R_0 = 1500~\Omega$,
and $\Delta R = 500~\Omega$.
These values were chosen because they are typical for SV/MTJs.
We also assumed that $\vb{p}_2$ remained pinned in a specific direction during neuron operation.

\begin{figure}
\includegraphics{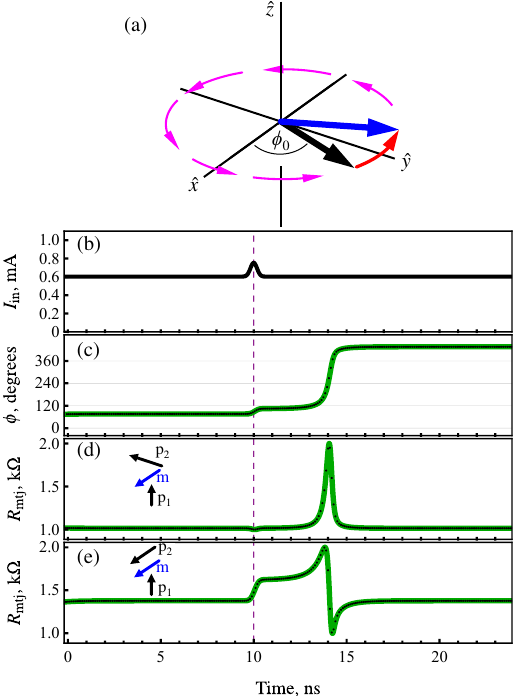}
\caption{\label{sims} Simulation results illustrating the magnetization dynamics and output resistance of the SV/MTJ neuron.
(a) Rotation of the free layer magnetization layer in the x-y plane.
The black arrow represents the steady-state magnetization angle $\phi_0 = 73^\circ$, the blue arrow represents the magnetization after impulse with a phase angle of $111^\circ$, the red arrow indicates the temporary shift due to a brief increase in input current, and the magenta arrows depict the full $360^\circ$ rotation of the magnetization.
(b) Input current $I_\text{in}$ for the system
(c) Evolution of $\phi$, the magnetization azimuthal angle, with time. Green curves are results from LLGS simulations, while gray dashed lines are results of ANE simulations. 
(d) Change in $R_\text{mtj}$ with time for $\vb{p}_2 = \vu{y}$. (Inset) Equilibrium magnetization directions.
(d) Change in $R_\text{mtj}$ with time for $\vb{p}_2 = \vu{x}$. (Inset) Equilibrium magnetization directions.}
\end{figure}

\section{DERIVATION AND SIMULATION OF ARTIFICIAL NEURON EQUATION}\label{wash}

In this section, it will be shown that the ANE can be derived from (\ref{llgs}).
This derivation begins by performing a scalar product between $\pdv{m}{\theta}$, $\pdv{m}{\phi}$ and $\dv{m}{t}$, yielding a pair of equations:
\begin{subequations}
\begin{eqnarray}
\dv{\theta}{t}  =& \omega_B \sin \phi + \alpha_G \cos \theta \displaystyle\dv{\phi}{t} - \sigma_j I_\text{in} \cos \theta ,\label{systemA}\\
\dv{\phi}{t}   =& -\omega_M \sin \theta - \omega_B\tan\theta\cos\phi - \alpha_G\displaystyle\frac{1}{\cos\theta}\dv{\theta}{t} . \label{systemB}
\end{eqnarray}
\end{subequations}
In these equations, $\phi$ is the azimuthal angle of $\vb{m}$, $\theta$ is the out-of-plane angle of $\vb{m}$, $\omega_B = |\gamma| B_\text{ext}$, and $\omega_M = |\gamma|B_d $.
Please see the supplementary material for the connection between (1) and (2).

We assume that during the operation of the SV/MTJ neuron, the magnetization of the free layer stays almost in-plane, which means we can use the small angle approximation for $\theta$.
Using this approximation, substituting equation (\ref{systemA}) into equation (\ref{systemB}), and ignoring the nonlinear terms, leads to the following equation:
\begin{equation} \dv{\phi}{t} = -\omega_M\theta - \omega_B\theta\cos\phi -\alpha_G \omega_B\sin\phi \label{tempB}.\end{equation}
This equation allows us to solve for $\theta$:
\[\theta = -\frac{\dv{\phi}{t} + \alpha_G \omega_B \sin\phi }{ \omega_M}, \]
where we have neglected the $\omega_B\cos\phi$ term in the denominator since $\omega_M$ is much larger than $\omega_B$.
After taking the time derivative of $\theta$, we can equate it with equation (\ref{systemA}).
From this, we can directly solve for the artificial neuron equation:
\begin{equation} \frac{1}{\omega_M}\dv[2]{\phi}{t} + \alpha_G \dv{\phi}{t} + \omega_B \sin \phi = \sigma_j I_\text{in} \label{ane}.\end{equation}
Similar artificial neuron equations have been derived in previous research for various spintronic devices, including antiferromagnetic neurons {[}Khymyn 2018{]}, synthetic antiferromagnetic oscillators {[Liu 2020{]}, optically initialized antiferromagnetic-heavy metal heterostructures {[}Mitrofanova 2022{]}, a spin Hall nano-oscillator {[}Ovcharov 2022{]}, and spin superfluid Josephson oscillators {[}Crotty 2010, Schneider 2018{]}.

%For this system, there is a threshold current, $I_\text{th}$, as in spin-torque-nano-oscillators {[}Slavin 2009{]}. 
%When the input current $I_\text{in} > I_\text{th}$, the free layer magnetization will process about $\vb{B}_\text{eff}$ {[}Houssameddine 2007{]}. 
%For the geometry considered here, there will be an out-of-plane rotation for steady-state currents above $I_\text{th}$ {[}Houssameddine 2007{]}. 
For this system, there is a threshold current, $I_\text{th}$, as in spin-torque-nano-oscillators {[}Slavin 2009{]}. 
When the input current $I_\text{in} > I_\text{th}$, the free layer magnetization processes around $\vb{B}\text{eff}$. In the geometry considered here, steady-state currents above $I\text{th}$ cause an out-of-plane rotation {[}Houssameddine 2007{]}.

When there is no input current ($I_\text{in} = 0$), $\vb{m}$ will be aligned with $\vb{B}_\text{ext}$, along the $\vu{x}$ direction.
When the input current is positive, but $I_\text{in} \leq I_\text{th}$, the free layer magnetization will remain stationary in the $\vu{x}-\vu{y}$ plane with a phase angle $\phi$ less than the threshold phase of $90^\circ$.
When the phase is $\phi = 90^\circ$, the threshold current can be calculated from (\ref{ane}), and is
\begin{equation}\I_\text{th} = \frac{\omega_B}{\sigma_j}. \end{equation}
%With the parameters considered above, the threshold voltage is $I_\text{th} \sim 0.622$ mA.
With the parameters above, the threshold voltage is $I_\text{th} \sim 0.622$ mA.

%The magnetization dynamics of the system are illustrated in Fig. \ref{sims} through the simulation of equation (\ref{llgs}).
%Additionally, we demonstrate that equation (\ref{ane}) provides a reasonable approximation of the LLGS for this device geometry.
The magnetization dynamics are shown in Fig. \ref{sims} through the simulation of equation (\ref{llgs}). 
We also show that equation (\ref{ane}) reasonably approximates the LLGS for this device geometry.

In this simulation, the input current consists of a bias DC current, $I_\text{bias}$, and a temporary impulse current, $i_p$, so that $I_\text{in} = I_\text{bias} + i_p$.
For this setup, $I_\text{bias}$ is set to 0.6 mA.
It is important to note that this is a sub-threshold current, meaning $I_\text{in} < I_\text{th}$.
At this current level, spin transfer torque causes the free layer magnetization to rotate in the $\vu{x}-\vu{y}$ plane, forming an angle $\phi_0 = 73^\circ$ with the $\vu{x}$ axis.
This is depicted in Fig. \ref{sims}(a), where the black arrow represents the free layer magnetization under the influence of a constant 0.6 mA current.
The same behavior is shown in Fig. \ref{sims}(b) and Fig. \ref{sims}(c) during the first 9 ns of the simulation.
In Fig. \ref{sims}(b), the black line represents the input current amplitude, while the green lines in Fig. \ref{sims}(c), (d), and (e) show the results from simulating equation (\ref{llgs}).
The black dots in Fig. \ref{sims}(c)-(e) represent the results from simulating equation (\ref{ane}).

At time $t = 10$ ns in the simulation, the input current $I_\text{in}$ briefly increases to approximately 0.75 mA, as shown in Fig. \ref{sims}(b).
In response to this perturbation, the free layer magnetization is temporarily pushed from $\phi_0$ past $90^\circ$ to $111^\circ$, as indicated by the red arrow in Fig. \ref{sims}(a).
Following a delay of about 4 ns, the free layer magnetization undergoes a full 360° rotation in the $x-y$ plane before returning to $\phi_0$.
This rotation is depicted by the magenta arrows in Fig. \ref{sims}(a) and by the green line in Fig. \ref{sims}(c) around $t = 14$ ns, when the phase angle increases to $433^\circ$.

The output resistance $R_\text{mtj}$, with $\vb{p}_2$ pinned in the $-\vu{y}$ direction, is shown in Fig. \ref{sims}(d).
With this analyzer configuration, the resistance of the structure momentarily increases from 1 k$\Omega$ to 2 k$\Omega$, showing a spiking response characteristic of previous studies {[}Bradley 2023{]}.
Alternatively, if $\vb{p}_2$ is pinned in the $\vu{x}$ direction, the output resistance changes as shown in Fig. \ref{sims}(e).
It is important to note that Fig. \ref{sims}(c), which displays the phase of the free layer magnetization, remains unchanged for both orientations of $\vb{p}_2$. 
This is because $\phi$ is directly influenced by the ANE, and represents the neuron dynamic behavior. 
In contrast, the outputs in Figs. \ref{sims}(d) and \ref{sims}(e) primarily reflect changes in the MTJ resistance.

The duration of the spiking response in Fig. \ref{sims}(d) is on the order of 1 ns. 
This is substantially faster than the response speed of a biological neuron ($\sim 1$~ms), but slower than the speed of an anitferromagnetic neuron ($\sim 1$~ps). 

It is important to note that Figs. \ref{sims}(c)-(e) display results from two different simulations.
The black dots represent the simulation using the ANE in equation (\ref{ane}), while the green line represents the simulation of the LLGS equation in equation (\ref{llgs}).
From these results, it is clear that, for this set of simulation parameters, the ANE provides a valid approximation of the LLGS.

\begin{figure}
\includegraphics{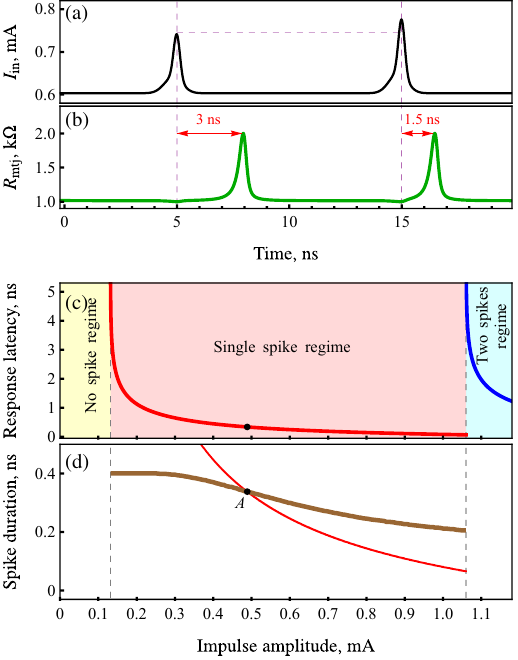}
\caption{\label{conB} 
    (a) Simulated input current with a bias of approximately 0.6 mA, including two current impulses at $t=5$ ns and $t=15$ ns. 
    (b) Corresponding output response $R_\text{mtj}$ of the SV/MTJ, showing variable response latency of 3 ns for the first impulse and 1.5 ns for the second, larger impulse. 
    (c) Simulation results for response latency across impulse amplitudes ranging from 0 to 1.2 mA, with regions indicating no spike (yellow), single spike (red), and two spikes (blue). 
    (d) Variation in spike duration (brown line) and response latency (red curve), and point `A' where latency and spike duration intersect.}
\end{figure}

Building on these simulation results, it is important to consider the impact of thermal noise on system performance. 
The Johnson-Nyquist thermal noise for a 10 GHz bandwidth is approximately 0.5 $\mu$A. 
The signal pulse in Fig. \ref{sims}(b) is 25\% higher than the bias current, providing a substantial margin over expected thermal noise. 
For the parameters in this paper, this provides a significant margin above the noise level. 
However, for SV/MTJ configurations with different parameters, thermal noise will require special consideration.

\section{SIMULATION OF SV/MTJ NEURON CHARACTERISTICS}

In the previous section, we showed that SV/MTJ behavior can be effectively modeled using ANE. 
This suggests that characteristics established for antiferromagnetic neurons, such as response latency, refraction, inhibition, spiking modes, synaptic integration, and adherence to the ``all-or-nothing'' law, also apply to SV/MTJ neurons {[}Bradley 2023{]}.

In biological neural networks, varying response latency based on input amplitude is a fundamental mechanism for learning.
The variable response latency of SV/MTJ, as determined from simulating (\ref{llgs}) with $\vb{p}_2 = -\vu{y}$ and $I_\text{th}=0.6$ mA, is illustrated in Fig. \ref{conB}(a) and (b).
Fig. \ref{conB}(a) shows a simulated input current with a bias of approximately 0.6 mA, along with two current impulses: the first at $t=5$ ns and the second at $t=15$ ns.
Note that the second impulse is larger than the first.
The corresponding output response, represented by $R_\text{mtj}$ of the SV/MTJ, is depicted in Fig. \ref{conB}(b).
In Fig. \ref{conB}(b), the response latency for the first, smaller impulse is 3 ns, while the response latency for the second, larger impulse is 1.5 ns.
This demonstrates that the response latency shortens as the input current impulse amplitude increases, as previously seen for antiferromagnetic neurons {[}Bradley 2023{]}.

To further investigate the response latency in this system, a series of simulations were conducted with impulse amplitudes ($i_p$) ranging from 0 to 1.2 mA.
The simulation results are presented in Fig. \ref{conB}(c).
As illustrated by the yellow region, the SV/MTJ neuron does not spike for impulse amplitudes below 0.15 mA.
For impulse amplitudes between 0.15 mA and 1.06 mA, the neuron generates a single spike, with the response latency depicted by the red curve in the red region.
For impulse amplitudes exceeding 1.06 mA, the neuron produces two spikes, with the inter-spike period indicated by the blue curve in the blue region.
%In the single spike regime, it is evident that smaller impulse amplitudes result in longer response latencies, whereas larger impulse amplitudes lead to shorter response latencies.
In the single spike regime, it is evident that smaller impulses cause longer response latencies, while larger ones lead to shorter latencies.

In addition to the change in response latency, it is important to note that the amplitude and spike duration of the SV/MTJ output resistance in Fig. \ref{conB}(b) remain nearly constant, even as the input amplitude varies.
The brown line in Fig. \ref{conB}(d) shows how the spike duration changes with $i_p$, while the red curve represents the response latency.
For longer response latencies, the spike duration stays relatively consistent.
However, when the impulse amplitude is strong, the neuron generates a spike almost immediately after the input, causing the response latency to shorten.
%When the response latency becomes shorter than the spike duration (as seen at point ‘A’ in Fig. \ref{conB}(d)), the spike duration decreases.
When the response latency is shorter than the spike duration (as at point ‘A’ in Fig. \ref{conB}(d)), the spike duration decreases.
In essence, the spike duration decreases when the neuron spikes immediately after the impulse.

Figure \ref{varyB}, which was also obtained through simulation, characterizes the response of the SV/MTJ neuron to bias currents that vary from 0 to 0.7 mA.
How the different spiking regimes vary with bias current is shown in Figure \ref{varyB}(a).
The region shaded yellow shows where the SV/MTJ neuron does not spike.
The red region shows where there is a single spike.
The blue region shows where there are multiple spikes.
As expected, the region above $I_\text{th}$ has multiple spikes without any impulse current.
Several other observations about this figure are notable.
First, higher impulse currents are required to generate a spike for lower bias currents.
A wider range of impulse currents can generate a spike for lower bias currents.
Most notably, according to these simulations, it is possible to induce spikes with zero bias current.

Figure \ref{varyB}(b) shows how the power consumption of the SV/MTJ neuron varies with changing bias current, assuming a time per synaptic operation is conservatively estimated to be 20 ns and a $\vb{p}_2$ that is aligned with $\phi_0$. 
In this figure, the green region represents the steady-state power consumption due to the bias current, while the red region indicates the dynamic power consumption, which includes both the impulse current and the changing resistance of the neuron.
From the figure, it is clear that the power consumption remains below 0.3 mW for high bias currents. 
For lower bias currents, the power consumption decreases to approximately 50 \unit{\uW}. 
%Additionally, while the dynamic power consumption increases with lower bias currents, it does not reach the levels observed at higher bias currents.
While dynamic power consumption increases with lower bias currents, it does not reach the levels seen at higher currents.

\begin{figure}
\includegraphics{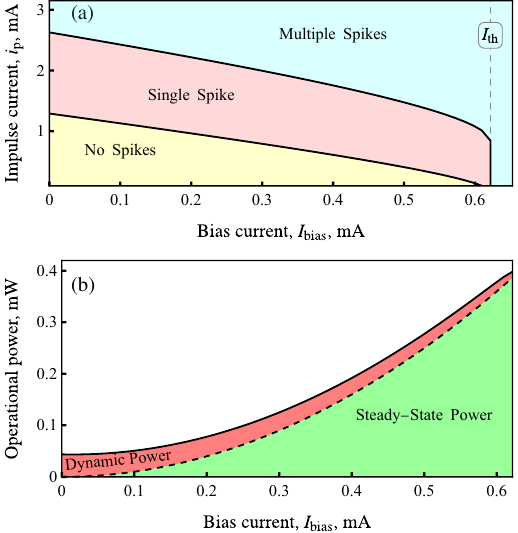}
\caption{\label{varyB}
Simulation results characterizing the response of the SV/MTJ neuron to varying bias currents. 
(a) The spiking regimes of the neuron. The yellow region indicates no spikes, the red region indicates a single spike, and the blue region indicates multiple spikes. 
$I_\text{th}$ indicated by a dashed line. 
(b) Power consumption of the neuron. 
The green region represents steady-state power consumption, while the red region shows dynamic power consumption. 
Assumes 20 ns per synaptic operation and $\vb{p}_2$ aligned with $\phi_0$. }
\end{figure}

\section{CONCLUSION}

In this paper, we introduced a novel approach to neuromorphic computing by developing an artificial neuron based on a combination of a spin valve and a magnetic tunnel junction. 
We analytically derived the artificial neuron equation from the Landau-Lifshitz-Gilbert-Slonczewski equation, and demonstrated through simulations that the ANE serves as a reliable approximation of the LLGS for this specific SV/MTJ design. 
Our simulation results further indicate that the proposed neuron can operate on a timescale of about 1 ns, without any bias current, achieving power consumption as low as 50 \unit{\uW}. 
These findings underscore the potential of SV/MTJ neurons to offer a low-power, scalable, and biologically plausible solution for neuromorphic computing. 
%Given that both spin valves and magnetic tunnel junctions are well-established, commercially available technologies that can be integrated with CMOS, they present a promising avenue for practical implementation. 
%This work lays the foundation for future research and development in spintronic neurons, which could significantly enhance the efficiency and capabilities of next-generation AI systems.
Since spin valves and magnetic tunnel junctions are well-established, commercially available, and CMOS-compatible, they offer a promising path for practical implementation. 
This work lays the groundwork for future research on spintronic neurons, which could greatly enhance next-generation AI efficiency and capabilities.

\section{ACKNOWLEDGEMENTS}

This work was partially supported by the Air Force Office of Scientific Research (AFOSR) Multidisciplinary Research Program of the University Research Initiative (MURI), under Grant No. FA9550-19-1-0307.

\pagebreak

%\onecolumn
\setcounter{equation}{0}
\renewcommand{\theequation}{S.\arabic{equation}}

%This supplementary material provides additional details and derivations to support the results presented in the main text of the article, ``Spintronic Neuron Using a Magnetic Tunnel Junction for Low-Power Neuromorphic Computing'' by Steven Louis, Hannah Bradley, Andrei Slavin, and Vasyl Tyberkevych. 
%In particular, following sections include a detailed derivation of the LLGS equation in spherical coordinates, as well as an analysis of the impact of the analyzer layer on the free layer magnetization. 

\section{Supplementary Material}

\noindent
This supplementary material provides additional details and derivations to support the results presented in the main text of the article, ``Spintronic Neuron Using a Magnetic Tunnel Junction for Low-Power Neuromorphic Computing'' by Steven Louis, Hannah Bradley, Cody Trevillian, Andrei Slavin, and Vasyl Tyberkevych. \\

The following sections offer a detailed derivation of the Landau-Lifshitz-Gilbert-Slonczewski (LLGS) equation in spherical coordinates, demonstrating its equivalence to the form used in the main text. 
Additionally, we analyze the influence of the analyzer layer spin-transfer torque on the free layer magnetization, explaining on the approximations discussed in the article.
These supplementary results provide a more comprehensive view of the theoretical framework and serve to justify the modeling choices made in the main text.

\subsection{LLGS Equation in Spherical Coordinates}

The purpose of this section is to show that equations (1) and (2) in the manuscript are equivalent. 
The coordinates used here are consistent with those presented in Fig. 1 of the main text. 
Note that $\theta$ represents the out-of-plane angle, not the polar angle.\\

Starting from equation (1), which describes the dynamics of the free layer using the Landau-Lifshitz-Gilbert-Slonczewski (LLGS) equation:
\begin{equation} 
\dv{\vb{m}}{t} = |\gamma|\vb{B}_\text{eff} \times \vb{m} + \alpha_G \vb{m} \times \dv{\vb{m}}{t} + \sigma_j I_\text{in} \vb{m} \times \vb{m} \times \vb{p}_1   . 
\label{fwfwefew} 
\end{equation}

In Cartesian coordinates, the vector unit magnetization is represented as $\vb{m}  = m_x \vu{x} + m_y \vu{y} + m_z \vu{z}$, where $|\vb{m}| = 1$. 
Substituting $m_x = \cos\theta\cos\phi$, $m_y = \cos\theta\sin\phi$, and $m_z = \sin\theta$ into $\vb{m}$ gives:
\begin{equation}
\vb{m}  = \cos{\theta}\cos\phi \vu{x} + \cos{\theta}\sin\phi \vu{y} + \sin\theta \vu{z}.
\end{equation}

After taking the partial derivatives of $\vb{m}$ with respect to $\theta$ and $\phi$, we get:
\begin{equation}
\frac{\partial \vb{m}}{\partial \theta} = -\sin\theta \cos\phi \vu{x} - \sin\theta \sin\phi \vu{y} + \cos\theta \vu{z},
\end{equation}
\begin{equation}
\frac{\partial \vb{m}}{\partial \phi} = -\cos\theta \sin\phi \vu{x} + \cos\theta \cos\phi \vu{y}.
\end{equation}

Thus, the total derivative of $\vb{m}$ is:
\begin{equation}
\dv{\vb{m}}{t} = \pdv{\vb{m}}{\theta}\dv{\theta}{t} + \pdv{\vb{m}}{\phi}\dv{\phi}{t}.
\end{equation}

Performing the calculation gives:
\begin{multline}
\dv{\vb{m}}{t} =  \qty[-\sin\theta \cos\phi \dv{\theta}{t}  -\cos\theta \sin\phi \dv{\phi}{t} ]\vu{x} \\
+ \qty[- \sin\theta \sin\phi \dv{\theta}{t}  + \cos\theta \cos\phi \dv{\phi}{t} ]\vu{y} +
\cos\theta\dv{\theta}{t}\vu{z}.
\end{multline}

Both $\vb{m}$ and $\dv{\vb{m}}{t}$ can be written in spherical coordinates:
\begin{equation} 
\vb{m} = \vu{r} 
\end{equation}
\begin{equation}
\dv{\vb{m}}{t} =  -\dv{\theta}{t}\vu{\theta}     +  \dv{\phi}{t}\cos\theta  \vu{\phi}.   
\end{equation}

\pagebreak

In addition, $|\gamma|\vb{B}_\text{eff}$ and $\vb{p}_1$ can be expressed in spherical coordinates:
\begin{equation}
|\gamma|\vb{B}_\text{eff} = |\gamma|\vb{B}_\text{ext}\vu{x} - |\gamma|B_d \vu{z}(\vu{z}\cdot\vb{m}) = \omega_B\vu{x} - \omega_M \sin\theta \vu{z},
\end{equation}
\begin{multline}
|\gamma| \vb{B}_\text{eff} = \qty(\omega_B \cos\theta \cos\phi - \omega_M \sin\theta^2 )\vu{\rho} \\+ 
 \qty(\omega_B \sin\theta \cos\phi + \omega_M \sin\theta \cos\theta ) \vu{\theta} + (-\omega_B \sin\phi )\vu{\phi},
\end{multline}
\begin{equation}
\vb{p}_1 = \vu{z} =   \sin\theta \vu{r}   - \cos\theta \vu{\theta}.
\end{equation}

\vspace{5 mm}
With this, it is possible to expand the terms on the right-hand side of equation (S.1):
\begin{equation} 
|\gamma| \vb{B}_\text{eff} \times \vb{m} = (-\omega_B \sin\phi  )\vu{\theta} + \qty(-\omega_B \sin\theta \cos\phi - \omega_M \sin\theta \cos\theta )\vu{\phi}, 
\end{equation}
\begin{equation} 
\alpha \vb{m} \times \dv{\vb{m}}{t} = \qty(- \alpha \cos\theta \dv{\phi}{t})\vu{\theta} +\qty(-\alpha  \dv{\theta}{t})\vu{\phi},  
\end{equation}
\begin{equation} 
\sigma_jI_\text{in}  \vb{m} \times \vb{m} \times \vb{p}_1 = \qty(\sigma_jI_\text{in} \cos\theta ) \vu{\theta}.  
\end{equation}

\vspace{5 mm}
Substituting equations (S.8), (S.4), (S.5), and (S.6) into equation (S.1) results in:
\begin{multline}
-\dv{\theta}{t}\vu{\theta} + \dv{\phi}{t}\cos\theta \vu{\phi} = -\omega_B \sin\phi \vu{\theta} - (\omega_B \sin\theta \cos\phi + \omega_M \sin\theta \cos\theta)\vu{\phi}\\ - \alpha \cos\theta \dv{\phi}{t}\vu{\theta} - \alpha  \dv{\theta}{t}\vu{\phi} + \sigma_jI_\text{in} \cos\theta \vu{\theta}.
\end{multline}

By separating this into two equations for $\vu{\theta}$ and $\vu{\phi}$ and simplifying, we obtain:
\begin{equation}
\dv{\theta}{t} = \omega_B \sin\phi + \alpha \cos\theta \dv{\phi}{t} - \sigma_jI_\text{in} \cos\theta,
\end{equation}
\begin{equation}
\dv{\phi}{t} = -\omega_M \sin\theta - \omega_B \tan\theta \cos\phi - \alpha \frac{1}{\cos\theta} \dv{\theta}{t},
\end{equation}
which matches equation (2) in the main text.

\subsection{LLGS Simulation Results Including the Analyzer Layer}

In this section, we provide additional details regarding the inclusion of the spin-transfer torque (STT) from the analyzer layer $\vb{p}_2$ in the modeling of the spintronic system. 
The dynamics of the free layer magnetization, $\vb{m}$, were modeled using the Landau-Lifshitz-Gilbert-Slonczewski (LLGS) equation presented in equation (1) of the main text. 
To further refine this model, the LLGS equation is expanded to incorporate the contribution of the spin-transfer torque due to $\vb{p}_2$. 
The revised equation is as follows:
\begin{equation}\tag{S.18}
\dv{\vb{m}}{t} = |\gamma|\vb{B}_\text{eff} \times \vb{m} + \alpha_G \vb{m} \times \dv{\vb{m}}{t} + \sigma_j I_\text{in} \vb{m} \times \vb{m} \times \vb{p}_1 + \sigma_j I_\text{in} \vb{m} \times \vb{m} \times \vb{p}_2.
\label{llgsFull}
\end{equation}
Here, $\vb{p}_1$ represents the magnetization of the polarizer layer, and $\vb{p}_2$ represents the magnetization of the analyzer layer. 
While the main text focuses on the contribution of $\vb{p}_1$, the influence of the analyzer layer via $\vb{p}_2$ is not included in equation (1) due to its relatively minor effect on system dynamics. 
This supplementary section further explores the impact of $\vb{p}_2$ and justifies its exclusion from the primary analysis.\\

In certain situations, the current flowing through the analyzer layer would indeed exert a torque on the free layer unit magnetization, $\vb{m}$. 
In this setup, $\vb{p}_1$ is oriented along the z-axis, while $\vb{m}$ is oriented in the xy-plane at equilibrium. 
This configuration ensures that the STT generated by $\vb{p}_1$ pulls $\vb{m}$ out of the xy-plane, creating out-of-plane precession. 
This precession is important for the neuron-like behavior described in the main text.\\

In contrast, $\vb{p}_2$ is oriented in the xy-plane. 
As a result, the STT contribution from $\vb{p}_2$ primarily affects the magnetization within the plane, creating an effective weak in-plane anisotropy. 
Consequently, this effect does not drive out-of-plane precession, and the influence of $\vb{p}_2$ does not qualitatively change the dynamics of the system.\\

To quantify the change due to $\vb{p}_2$, several of simulations was performed that included $\vb{p}_2$ in the LLGS equation. 
The results of these simulations are presented in Fig S1. 
The blue curve represents the phase $\phi$ of $\vb{m}$ when $\vb{p}_2$ is excluded from the equation, serving as a baseline. 
The black curve shows the phase when $\vb{p}_2$ is aligned in the x-direction, and the green curve shows the phase when $\vb{p}_2$ is oriented in the y-direction.\\

\renewcommand{\thefigure}{S\arabic{figure}}

\setcounter{figure}{0}

\begin{figure}[h!]\begin{center}
\includegraphics[width=3.5 in]{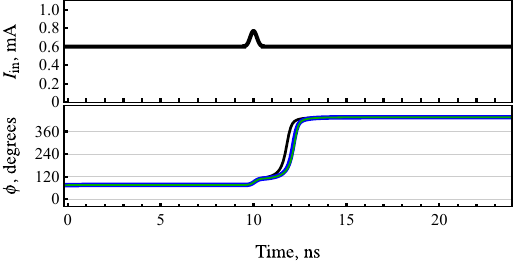}
\caption{   Comparison of the phase $\phi$ of $\vb{m}$ in simulations with and without $\vb{p}_2$ in the LLGS equation. The blue curve represents the baseline phase without $\vb{p}_2$, while the dashed black and green curves correspond to the phase when $\vb{p}_2$ is aligned in the x- and y-directions, respectively.}
\end{center}\end{figure}

The results demonstrate that when $\vb{p}_2$ is aligned with the equilibrium position of $\vb{m}$ (i.e., in the x-direction), the dynamics of $\vb{m}$ remain unchanged compared to the baseline. 
However, when $\vb{p}_2$ is oriented perpendicular to $\vb{m}$ (in the y-direction), a slight shift in the phase of $\vb{m}$ is observed, indicating a minor quantitative change in behavior. 
Importantly, this shift does not significantly alter the overall dynamics, confirming that the contribution of $\vb{p}_2$ is indeed negligible in the context of system operation.\\

Based on these simulation results, it is clear that when $\vb{p}_2$ is oriented in-plane, it can generate minor modifications to system behavior. 
These modifications are small enough that they do not affect the neuron spiking behavior or the key characteristics focused on in the main text. 
Therefore, for clarity and simplicity, $\vb{p}_2$ was excluded from the primary LLGS equation used in the analysis. However, for completeness, this expanded discussion and corresponding simulation results are included in the supplementary material.

\end{document}